\documentclass [prl,amsmath,showpacs,twocolumn,preprintnumbers,superscriptaddress]{revtex4-1}

\usepackage{mathptmx}
\usepackage{microtype}
\usepackage[latin1]{inputenc}
\usepackage[spanish]{babel}
\usepackage{graphicx}
\usepackage{dcolumn}
\usepackage{amsmath}
\usepackage{mathtools}
\usepackage{bm}
\usepackage{amssymb}

\begin{document}

\title{How  is affected the thermodynamic description of physical systems when gravity is incorporated?}

\author{W. A. Rojas C.}
\email{warojasc@unal.edu.co}

\affiliation{Universidad Nacional de Colombia}

\author{R.Arenas S.}
\email{jrarenass@unal.edu.co}
\affiliation{Universidad Nacional de Colombia}
 
\begin{abstract}

This research is aimed at the revision of the  thermodynamic concepts and their relationship with the General Relativity Theory (GRT) and how physical systems are affected when gravity is included in their thermodynamic description. We  found that in the case of light, the entropy in an extreme stage is proportional to the area. This is due to a reduction of freedom degrees of the system, since gravity imposes  constraint on the number of microstates which are accessible to the system and this is consistent with the holographic principle.

\end {abstract}


\maketitle
\section{Introduction}
Standard Thermodynamics is the physical field  that study the processes of energy exchange between the systems and their surroundings. Such a description uses thermodynamic variables to characterize the state of a physical system but the, do not include gravity.  However is interesting to know how is affected the thermodynamica description when gravity is included, in particular if light is considered.
\section{THE CONNECTION BETWEEN GRAVITY AND THERMODYNAMICS}
Newton discoverded the fundamental laws to predict the movement of the heavenly bodies.Wich corresponds to a law that is directly proportional to the product of the masses of interacting bodies and inversely proportional to square of the separation distance between them
\begin{equation}
\vec{F}=G\frac{m_{1} m_{2}}{r^{2}}\hat{r},
\end{equation}
where $ G $ is the universal gravitational constant. This description is aceptable for the most of the  mechanical phenomena. Was not until the early twentieth century when Einstein explained to us the reasons why gravity acts by his Field Equations
\begin{equation}
G_{ab}=\frac{8\pi G}{c^{4}}T_{ab},
\end{equation}
where $ G_ {ab} $ is the Einstein tensor, $ c $ the speed of light and $ T_ {ab} $ the energy momentum tensor. Equations (2) connect gravity in terms of the $G_{ab}$ and the matter-nergy in terms of $T_{ab}$.

In the thermodynamic description of physical systems are noted the following limitations
\begin{itemize}
	\item 	All systems analyzed are considered at rest relative to an observer. Are not studied accelerated systems.
	\item Such descriptions do not take into account thermal gravitational effects.
\end{itemize}
Such restrictions on the thermal description systems should be removed by considering the effects of space-time curvature for scenarios where these effects are not negligible\cite{Tolman}. One way that historically has served to link  GRT to classical thermodynamics has been to consider the first law of the thermodynamics
\begin{equation}
\Delta E=\Delta Q- \Delta W,
\end{equation} 
which refers to the conservation of energy of any physical system. This law shows the energy transfer mechanism between the system and surroundings, either by heat $ Q $ or work $ W $. If we consider that the relativistic equivalent of the first law of thermodynamics can be obtained via the energy-momentum tensor $ T_ {ab} $, since such tensor includes all forms of energy and matter present. Them we may  express the conservation of energy in GRT by introducing the covariant derivate of energy-momentum tensor\cite{Tolman}.
\begin{equation}
\nabla_{a}T^{ab}=0.
\end{equation}
The  Zeroth law  declares that there is a balance parameter we call temperature $ T $. If a physical system such as an ideal gas is in thermal equilibrium, where in all parts exhibit the same temperature. In the gravitational context, we must make the distinction between two types of temperature measurement, one local and another one at in infinity. Consider a space-time of Schwarzschild type, a black hole of mass $ M $ and spherically symmetric non-rotating without charge. The local temperature is a function that depends only on the radial distance $r$, i.e. it varies with the gravitational potential. 	Supposing, that a gas of particles is in a strong gravitational field, its temperature is affected by the space-time curvature (This phenomenon is known as Tolman Law: $T(r)=T_{\infty}f(r)^{-1/2}$).

 An observer, who is far from the black hole horizon measures a temperature $T_{\infty} $. The temperature increases when is associated to an object falling  radially  towards the black hole. The object has been thermalized for the distant observer, then he concludes that the black hole acts as a heat source\cite{Tolman, SusskindL}.

A object with a different temperature of absolute zero has a certain degree of disorder which we call entropy. It is a state function that serves to characterize a physical system, such an ideal gas in a container of adiabatic walls. Its entropy is proportional to the volume of the container (in full agreement with the Boltzmann principle, which establishes the relationship between entropy $S$ and the logarithm of the probability to find the system in a certain microstate $\Omega$. Boltzmann postulate that $S=k_{B}ln\left|\Omega\right|$ with $k_{B} $ equal to the Boltzmann constant):
\begin{equation}
S\propto ln\left|\frac{V}{V_{0}} \right|^{N},
\end{equation}
where $ V_{ 0} $ and $ V$ are the initial and final volumes  and  $N$ the number of particles. Such entropy  depends on the number of degrees of freedom per particle, which for a monatomic ideal gas is equal to $\frac{3}{2}k_{B}T$.

This allows to characterize the number of microstates consistent with a macrostate, which eventually corresponds to the information system could be found because a higher entropy, there is less information available about the system state. Statistical entropy must account for the information we have about the state of the system and this can never decrease in time at least for an isolated system must remain constant \cite{SusskindL,Benkenstein}.

In the early 70's, some thermodynamic properties of black holes were well establised by the work of Hawking and Bekenstein  \cite{Hawking, Benkenstein}. Since then it was clear that the area of the event horizont behaves as entropy and surface gravity as temperature. Thus, the thermodynamic entropy of black holes is proportional to the areas of the horizon \cite{Corichi}. Understand the foundations of this kind of entropy is one  the paradigms of current theoretical physics. Some explanations resort to sophisticated theories as String Theory and Loop Quantum Gravity. 

In order to go  deep into thermodynamics under gravitational effets fron GRT, we studied electromagnetic radiation in extreme  gravitational stage.


\section{Einstein's method}
Maxwell showed that light has  wave nature and one time later Planck and Einstein showed that  electromagnetic radiation also has granular structure, indivisible quanta of radiation are emitted and absorbed continuously when light interacts with matter \cite{Einstein}.

 One of the advantages to be showed by einstein method, was that from the distribution function of Wien's black body and the  Boltzmann's principle, grain structure of light in Minkowski space  was set, without additional assumptions like Planck did.

We follow the Eintein's method in order to determine whether light in space-time with curvature has corpuscular structure.

According to the second law of thermodynamics, we assume the light as a physical system in a definite state entropy density $S=V\phi$, where $V$ is the volume of the physical system and $\phi$ the entropy density. Such a entropy is the sum of the monochromatic entropies (i.e. with a associated frecuency)  which are spaced from each other and which can be obtained by adding
	\[S=\int^{\infty}_{0} V \phi d \nu.
\]
It is true in flat space. In curved space-time, gravity effects on the physical system volume must be considered. Let $dV=\frac{4\pi r^{2}}{\sqrt{f(r)dr}}$ be the differential volume element corrected gravitationally, with $\rho(\nu)$ being the black body distribution for a curved space-time. So that the total entropy of the electromagnetic radiation is
\begin {equation}
S=\int^{R}_{0}\int^{\infty}_{0} \phi \left(\rho(\nu),\nu \right)d\nu \frac{4\pi r^{2}}{\sqrt{f(r)}} dr.
\end{equation}
For a model black body type, $\Delta S=0$, we obtain the law
\begin{equation}
\frac{\partial \phi }{\partial \rho}=\frac{1}{T_{\infty}}.
\end{equation}
This means that all radiation with different frequencies are characterized by having the same temperature and that this is affected gravitationally
\begin{equation}
	\frac{\partial \phi}{ \partial \rho}=\frac{1}{T(r)}=\frac{1}{T_{\infty}}f(r)^{1/2}.
\end{equation}
We know that light frequency is also affected for gravitational field. So we can write the Wien distribution function 
\begin{equation}
\rho(\nu ,r)=\frac{8\pi h (\nu_{\infty}f(r)^{-1/2})^{3}}{c^{3}}e^{-\frac{h\nu_{\infty}}{k_{B}T_{\infty}}}.
\end{equation}
from of (9) we calculated $\frac{1}{T_{\ infty}} $ and then we replace into (8)
\begin{equation}
\frac{d\phi}{d\rho}=-\frac{k_{B}}{h\nu_{\infty}} ln\left| \frac{\rho c^{3}}{8\pi h \nu_{\infty}^{3} f(r)^{-3/2}}\right|f(r)^{1/2}.
\end{equation}
Integrating
\begin{equation}
\phi=-\frac{k_{B}f(r)^{1/2}\rho}{h\nu_{\infty}} \left[ ln\left| \frac{\rho c^{3}f(r)^{3/2} }{8\pi h \nu_{\infty}^{3}  }\right| -1\right].
\end{equation}
We have entropy in a frequency range $\nu$ and $\nu+d\nu$ is given by
\begin{equation}
S=V\phi \Delta \nu,
\end{equation}
and energy per unit volume and frequency in the form
\begin{equation}
E=V \rho \Delta \nu.
\end{equation}
Therefore, we have (11) becomes
\begin{equation}
S=-\frac{k_{B}f(r)^{1/2} E}{h\nu_{\infty}} \left[ ln\left| \frac{ c^{3}f(r)^{3/2} E}{8\pi h \nu_{\infty}^{3} V \Delta \nu }\right| -1\right].
\end{equation}
Let  $S_{0}$ be electromagnetic radiation entropy confined to a volume $V_{0}$
\begin{equation}
S_{0}=-\frac{k_{B}f(r)^{1/2} E}{h\nu_{\infty}} \left[ ln\left| \frac{ c^{3}f(r)^{3/2} E}{8\pi h \nu_{\infty}^{3} V_{0} \Delta \nu }\right| -1\right].
\end{equation}
The change in entropy $\Delta S$   in a volume $V_{0}$ to $V$ volume for the system is
\begin{equation}
\Delta S=-\frac{k_{B}f(r)^{1/2} E}{h\nu_{\infty}} ln\left| \frac{V}{V_{0}}\right|,
\end{equation}
if the Boltzmann principle is always considered valid with gravity. such that the entropy is proportional to logartimo of the probability of finding the system in a microstate given $th\Delta S = k_{B}ln\left|\Omega\right|$. And that such a probability for an ideal gas is $\Omega=\left[\frac{V}{V_{0}}\right]^{N}$, and $N$ is the number of gas molecules. Then (16) can be written as
\begin{equation}
\Delta S=k_{B} ln\left| \frac{V}{V_{0}}\right|^{\frac{E f(r)^{1/2}}{ h\nu_{\infty}}}.
\end{equation}
Einstein showed
	\begin{equation}
\Delta S=k_{B} ln \left| \frac{V}{V_{0}}\right|^{\frac{E}{h\nu}}.
\end{equation}
Comparison between (17) and (18) gives
\begin{equation}	
\frac{E }{N}=h \nu_{\infty}f(r)^{-1/2}=h \nu(r).
\end{equation}
The energy photons is corrected gravitationally for a factor $f(r)^{-1/2}$. The  entropy for an classical ideal gas at constant temperature is of the form
\begin{equation}
pdV=TdS=nRT \frac{dV}{V}.
\end{equation}
In the limit when $\Delta S\rightarrow 0$, (18) is reduced to
\begin{equation}
dS=\frac{k_{B}E f(r)^{1/2}}{h\nu_{\infty}} \frac{dV}{V},
\end{equation}
then
\begin{equation}
T_{\infty}dS=\frac{k_{B}E f(r)^{1/2}}{h\nu_{\infty}} T_{\infty}\frac{dV}{V}.
\end{equation}
The comparison between the equations (20) and (22) allows us to obtain more evidence about the granular structure of electromagnetic radiation near the surface of Schwarzschild. In the Wien approximation, which works well on high-energy range, the light in a strong gravitational field behaves as an ideal gas with energy quanta $ hv (r) $. Einstein method has proved effective even in a scenario as gravitationally strong. The light exhibits a granular structure.

\section{AN APPROACH TO THERMODYNAMIC DESCRIPTION OF LIGHT IN A CURVED SPACE-TIME}
Let $m$ be a stellar mass body with spherical symmetry. Consider two reflective concentric spherical shells sorrounding this mass. Whose radii $R$ and $L$, accoding to Figure (1),  are larger than its gravitational radius $R_{0}$, such that $R=R_{0}+\epsilon$, with $\epsilon \ll R_{0}$ \cite{Rojas, Mukohyama}.

\begin{figure}[h]
	\centering
		\includegraphics[width=0.4\textwidth]{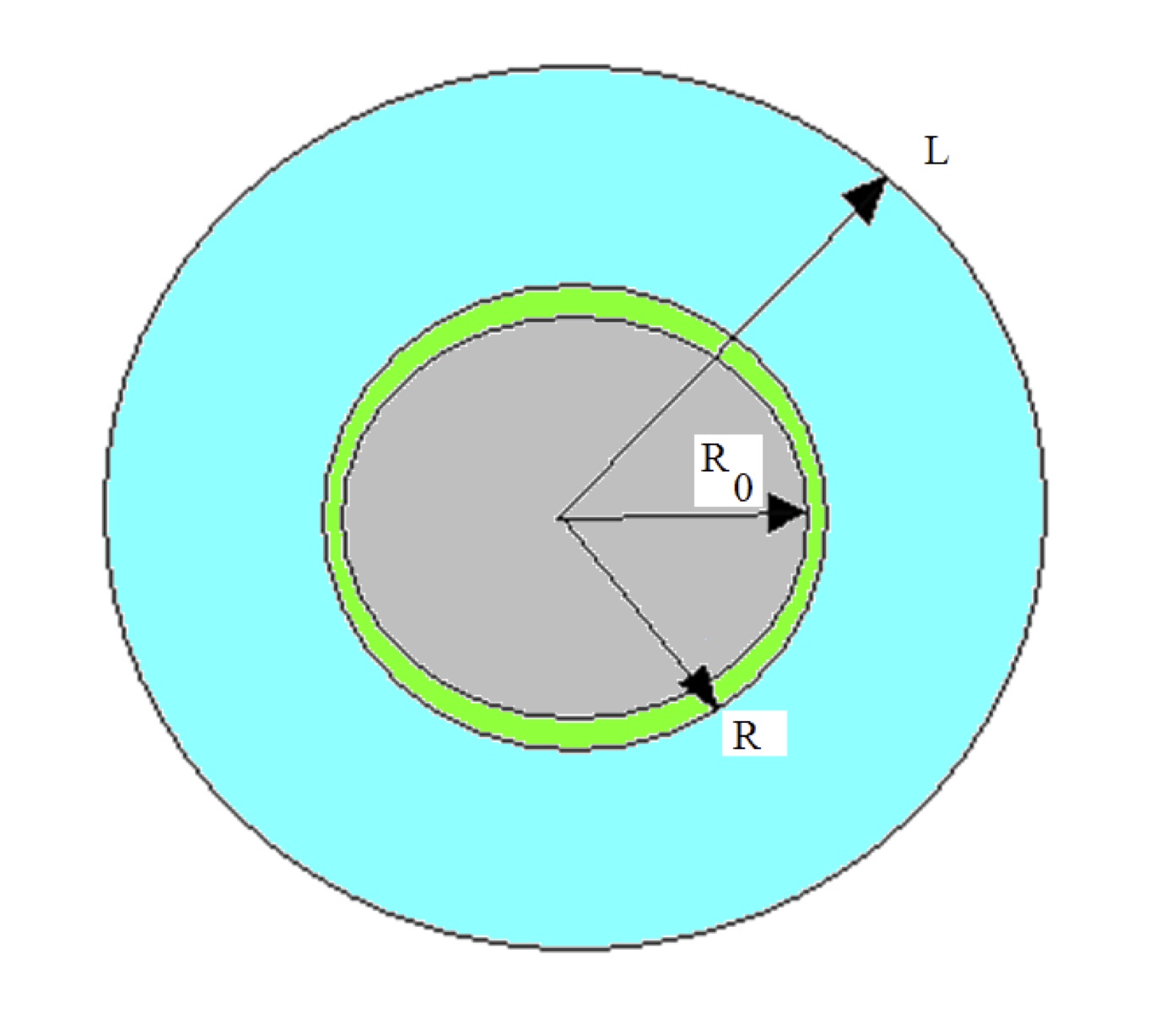}
		\caption{Gravitational body surrounded by two reflecting surfaces}
	\label{fig:triangulo}
\end{figure}
A photon gas is placed into space  bettween these tow reflecting surfaces, raised to some $T_{\infty}$ at large distances.

A particle description is good aproximation to the statistical thermodynamics of the fields, since for high energy radiation, wavelenghts are small compared to $R$ and $L$ \cite{ Mukohyama}.

The space between  two reflective surfaces is described by a metric of the form
 \begin{equation}
ds^{2}=-f(r)dt^{2}+f(r)^{-1}dr^{2}+d^{2}\Omega,
\end{equation}
and it corresponds to the metric which describes the outside of the mass. 

We can write the Helmholtz free energy near the inner surface as \cite{Fursaev}
\begin{equation}
F=-\frac{\pi^{2}k^{4}_{B}}{90 \hbar^{3}c^{3}} \int T^{4}\sqrt{-g} d^{3}x.
\end{equation}
Where $\sqrt{-g}$ is the determinant of the metric tensor $ g_ {ab} $ associated with the line element given by (23) and $ d^{3}x$ is the differential volume element. Near the horizon we can replace the coordinate $r$ by the coordinate $\zeta$\cite{Susskind}, which measures the proper distance from the Schwarzschild radius, $R_{0}=\frac{2Gm}{c^{2}}$. The Helmholtz free energy can be written as
\begin{equation}
F=-\frac{\pi^{2} k^{4}_{B}c^{3}}{90\hbar^{3}}T^{4}_{\infty}\kappa^{-3}\int d^{2}\sigma \int \zeta^{-3}d\zeta,
\end{equation}
where we denote by $\kappa$ the surface gravity and $d^{2}\sigma=dx^{2}+dy^{2}$. 

Integrating (25) with $d^{2}\sigma=A$
\begin{equation}
F=-\frac{\pi^{2} k^{4}_{B}c^{3}}{90\hbar^{3}}T^{4}_{\infty}\kappa^{-3}A\int ^{\zeta=\delta}_{\zeta=\epsilon}\zeta^{-3}d\zeta,
\end{equation}
with the approximation $\delta\gg\epsilon$, is
\begin{equation}
F=-\frac{\pi^{2} k^{4}_{B}c^{3}}{180\hbar^{3}\epsilon^{2}}T^{4}_{\infty}\kappa^{-3}A .
\end{equation}
The next step is calculating other thermal properties of light with the same recipe of flat space.  Entropy of light near the horizon is described by:
\begin{equation}
S=- \left(\frac{\partial F}{\partial T_{\infty}} \right)_{V}=\frac{\pi^{2} k^{4}_{B}c^{3}}{45\hbar^{3}\epsilon^{2}}T^{3}_{\infty}\kappa^{-3}A. 
\end{equation}
Note that it is still an extensive property  of the physical system, and now is proportional to the area $A$. Thsi does not mean that the  system has been  reduced to the area $A$ (Compared to (5)). 

The internal energy of the light is given by
\begin{equation}
E=\frac{\pi^{2} k^{4}_{B}c^{3}}{60\hbar^{3}\epsilon^{2}}T^{4}_{\infty}\kappa^{-3}A .
\end{equation}
The internal energy of the light is proportional to $T^{4}_{\infty}\kappa^{-3}A$. This confirms that, Stefan-Boltzmann law is also valid in the stage gravitacional ($E \propto T^{4}$). 

The heat capacity is obtained from
\begin{equation}
C_{v}=-\left( \frac{\partial E}{\partial T_{\infty}}\right)_{V}=\frac{\pi^{2} k^{4}_{B}c^{3} }{15\hbar ^{3}\epsilon^{2}}T^{3}_{\infty} \kappa^{-3}A. 
\end{equation}
If we consider a black hole of one solar mass, its Hawking temperature $T_{H}\propto10K^{-8} $, which corresponds to a temperature near absolute zero. Dittman and Zemanski  discussed the matter: \textsl{If  $T\rightarrow 0$, we will have $C_{P}\rightarrow C_{V} $} \cite{Zemansky}.

The pressure exerted by electromagnetic radiation is
	\begin{equation}
	P=-\frac{1}{\epsilon} \left(\frac{ \partial F }{\partial A} \right)_{T_{\infty}}=\frac{\pi^{2}k^{4}_{B}c^{3} }{180\hbar^{3}\epsilon^{3}}  T^{4}_{\infty} \kappa^{-3}A.
\end{equation}
The pressure exerted by electromagnetic radiation is proportional to $T_{\infty}^{4}\kappa^{-3}A$ and is strongly linked to the type of $\epsilon$ that we could choose.

\section{Conclusions}
The result of (19) indicates that the notion of photon proposed by Einstein, considering the Wien approximation for the distribution of black body remains valid even in the presence of gravity. Thus, the energy of photons in strong gravitational field is  $h\nu(r)$ including gravitational correction.  Likewise, the comparison between the expressions (18) and (19), allows us to show the grain structure of electromagnetic radiation near the gravitational radius. All this as long as Boltzmann Law is valid.

Near the horizon gravitational field is very strong if $\epsilon$ is small compared to the dimensions of the system due to $\nu(r)=\nu_{\infty}f(r)^{-1/2}$.

Helmholtz free energy is proportional to $T_{\infty}^{4}\kappa^{-3}A$, with $A$, being the area of the horizon, not proportional  to the volumen of the system, unlike the Minkowski space-time\cite{Landau}. This is important given the relationship between the Helmholtz energy $F$ and $Z$ the partition function of statistical thermodynamics 
\begin{equation}
F \propto ln\left|Z\right|.  
\end{equation}
Expression (28) corresponds to the entropy of electromagnetic radiation near the horizon  wich is proportional to the area.  This is justified by the fact that we have always considered valid the Boltzmann principle, which tells us that the entropy is proportional to the logarithm of the probability and linked to the set of microstates that are accessible to the system. Such a number of configurations is associated to the number of degrees of freedom of the system. In a strong gravitational field, the entropy of electromagnetic radiation exhibits a behavior proportional to the area. This means that the number of microstates that are accessible to the system decrease when gravity is incorporated in the thermodynamic description of the light. What is happening  to those microstates that are no available to system? In thermal equilibrium conditions all microstates are equiprobable in order to fulfill the condition of maximum entropy. When gravity is considered in the statistical description of light, certain microstates are not equiprobables and therefore they are no longer accessible to the system. This occurs because the number of degrees of freedom of the radiation has decreased. Gravity imposes a bond on the system. All this limiting its degrees of freedom and their microstates. 

Shows that the ordinary matter presents an entropy  proportional to area,  so thermodynamics is  consistent with the holographic principle when gravity is incorporated.

The holographic principle was first expressed by 't Hooft and Susskind in $1993$. and declares that:

\textsl{the maximum possible entropy depends on the area of the surface that delimits the volume and unlike to the volumen ... if a three dimensional system can be described for a physic theory, defined only in its bidimensional contour,  it is expected that the content of the information the system don't exceed the content of  the description limited to the contour}\cite{Bekensteinj}.
\section*{References}

\providecommand{\newblock}{}

\end{document}